\documentclass{article}
\usepackage{latexsym}
\usepackage{graphicx}
\begin{document}

\begin{center}
\textbf{\LARGE On the Solution of Maxwell's First \\ \vspace{.3cm} Order Equations}
\end{center}
\vspace{.1cm}
\begin{large}

\begin{center}
W. Engelhardt\footnote{Home address: Fasaneriestrasse 
8, D-80636 M\"{u}nchen, Germany\par Electronic address: 
wolfgangw.engelhardt@t-online.de}, retired from:
\end{center}

\begin{center}
Max-Planck-Institut f\"{u}r Plasmaphysik, D-85741 Garching, Germany
\end{center}

\vspace{.6cm}

\noindent \textbf{Abstract }

\noindent In an attempt to solve Maxwell's first order system of equations, starting 
from a given initial state, it is found that a consistent solution depending 
on the temporal evolution of the sources cannot be calculated. The well 
known retarded solutions of the second order equations, which are based on 
the introduction of potentials, turn out to be in disagreement with a direct 
solution of the first order system.
\vspace{.3cm} 
\vspace{.6cm}

\noindent \textbf{1. Introduction}

\noindent In recent papers [1, 2] it was shown that Maxwell's equations have different 
formal solutions depending on the chosen gauge. In [2] it was argued that 
the formalism of gauge invariance is based on the tacit assumption of 
Maxwell's equations having unique solutions which appeared, however, not to 
be guaranteed \textit{a priori. }In response to the publication of [2] it was pointed out in 
private communications [3] that uniqueness is a necessary consequence of the 
linear structure of the equations. These arguments are valid. If one finds, 
nevertheless, different solutions in Lorenz and in Coulomb gauge, it seems 
to indicate that a solution does not exist at all. Indeed, it was shown in 
[2] that the Li\'{e}nard-Wiechert fields based on the Lorenz gauge do not 
satisfy the equations in the source region, unless one postulates a velocity 
dependent ``deformation'' of point charges as in [1]. Furthermore, the 
formal solution for the vector potential in Coulomb gauge led to an 
undefined conditionally convergent integral which would even diverge upon 
differentiation. 

The reason for the difficulties encountered could have to do with the 
assumption of point sources which were exclusively considered in [2]. 
Therefore, it appears worthwhile to investigate the problem further, 
assuming smooth charge and current distributions as originally considered by 
Maxwell. In order to avoid any ambiguities arising from the introduction of 
potentials, it seems advisable to analyse directly the solvability of the 
first order system of Maxwell's equations (Sect. 2.). It turns out that the 
coupled first order system contains certain inconsistencies which prevent 
its solution when calculated by a numerical forward method proceeding in 
time.

The usual method of solution derives inhomogeneous wave equations from the 
first order system, and expresses the solutions as retarded integrals by 
application of Duhamel's principle. In [2] it was argued that this method is 
not plausible, since the wave equations obtained by differentiating the 
first order system connect the travelling fields with the stationary sources 
at the same time, while in the retarded solutions the differentiation of the 
sources is inconsistently dated back to an earlier time. In Sect. 3. we 
analyze the retarded solutions for smooth source distributions and find that 
these solutions do not satisfy the first order system. This is demonstrated 
in Sect. 4. by considering a specific example.
\vspace{.6cm}

\noindent \textbf{2. The first order equations}

\noindent In \textit{vacuo} the first order system as devised by Hertz on the basis of Maxwell's 
equations is supposed to describe the electromagnetic field:
\begin{equation}
\label{eq1}
div\,\vec {E}_g =4\pi \,\rho 
\end{equation}
\begin{equation}
\label{eq2}
rot\,\vec {E}_r =-\frac{1}{c}\frac{\partial \vec {B}}{\partial t}
\end{equation}
\begin{equation}
\label{eq3}
div\,\vec {B}=0
\end{equation}
\begin{equation}
\label{eq4}
rot\,\vec {B}=\frac{4\pi }{c}\,\vec {j}+\frac{1}{c}\frac{\partial \left( 
{\vec {E}_g +\vec {E}_r } \right)}{\partial t}
\end{equation}
Here we have indicated that the electric field has two contributions of 
different structure. In (\ref{eq1}) only the irrotational part enters, whereas (\ref{eq2}) 
contains exclusively the rotational part of the field. Both parts enter 
equation (\ref{eq4}). One may separate out the instantaneous contribution of the 
magnetic field and write (\ref{eq4}) as two equations:
\begin{equation}
\label{eq5}
rot\,\vec {B}_0 =\frac{4\pi }{c}\,\vec {j}+\frac{1}{c}\frac{\partial \vec 
{E}_g }{\partial t}
\end{equation}
\begin{equation}
\label{eq6}
rot\,\vec {B}_1 =\frac{1}{c}\frac{\partial \vec {E}_r }{\partial t}
\end{equation}
The quasi-static solutions of (\ref{eq1}) and (\ref{eq5}) -- subject to the boundary 
condition that the fields vanish at infinity -- are represented by integrals 
over all space:
\begin{equation}
\label{eq7}
\vec {E}_g \left( {\vec {x},\,t} \right)=\int\!\!\!\int\!\!\!\int {\rho 
\left( {\vec {x}',\,t} \right)} \,\left( {\vec {x}-\vec {x}'} 
\right)\frac{d^3x'}{\left| {\vec {x}-\vec {x}'} \right|^3}
\end{equation}
\begin{equation}
\label{eq8}
\vec {B}_0 \left( {\vec {x},\,t} \right)=\frac{1}{c}\int\!\!\!\int\!\!\!\int 
{\left( {\vec {j}\left( {\vec {x}',\,t} \right)+\frac{1}{4\pi 
}\frac{\partial \vec {E}_g \left( {\vec {x}',\,t} \right)}{\partial t}} 
\right)} \times \left( {\vec {x}-\vec {x}'} \right)\,\frac{d^3x'}{\left| 
{\vec {x}-\vec {x}'} \right|^3}
\end{equation}
It remains then to determine the rotational part of the electric field and 
the contribution $\vec {B}_1 $. 

Applying a numerical forward method one obtains from (\ref{eq6}) the difference 
equation:
\begin{equation}
\label{eq9}
\vec {E}_r \left( {\Delta t} \right)=\vec {E}_r \left( 0 \right)+\Delta 
t\,c\,rot\,\vec {B}_1 \left( 0 \right)
\end{equation}
and from (\ref{eq2}):
\begin{equation}
\label{eq10}
\vec {B}\left( {\Delta t} \right)=\vec {B}\left( 0 \right)-\Delta 
t\,c\,rot\,\vec {E}_r \left( 0 \right)
\end{equation}
Assuming that the sources were constant for $t\le 0$ one has the initial 
conditions:
\begin{equation}
\label{eq11}
\vec {E}_r \left( 0 \right)=\vec {B}_1 \left( 0 \right)=0
\end{equation}
Substituting this into (\ref{eq9}) and (\ref{eq10}) one finds the curious result that 
neither $\vec {E}_r $ nor the total magnetic field $\vec {B}$ proceed after 
the first time step, and this will remain so forever, at least in the vacuum 
region outside the sources. If the current would linearly rise to a new 
stationary level, e.g., equation (\ref{eq10}) would predict that $\vec {B}$ stays 
constant at its initial value, in contrast to (\ref{eq8}) which predicts that $\vec 
{B}_0 $ rises simultaneously with the current and reaches a new stationary 
value as well. 

One may also split (\ref{eq2}) into two equations:
\begin{equation}
\label{eq12}
rot\,\vec {E}_{r0} =-\frac{1}{c}\frac{\partial \vec {B}_0 }{\partial t}
\end{equation}
\begin{equation}
\label{eq13}
rot\,\vec {E}_{r1} =-\frac{1}{c}\frac{\partial \vec {B}_1 }{\partial t}
\end{equation}
The quasi-static solution of (\ref{eq12}) is:
\begin{equation}
\label{eq14}
\vec {E}_{r0} =-\frac{1}{4\pi \,c}\int\!\!\!\int\!\!\!\int {\frac{\partial 
\vec {B}_0 }{\partial t}\times \frac{\vec {x}-\vec {x}'}{\left| {\vec 
{x}-\vec {x}'} \right|^3}} \,d^3x'
\end{equation}
and from (\ref{eq13}) follows:
\begin{equation}
\label{eq15}
\vec {B}_1 \left( {\Delta t} \right)=\vec {B}_1 \left( 0 \right)-\Delta 
t\,c\,rot\,\vec {E}_{r1} \left( 0 \right)=0
\end{equation}
If $\vec {B}_1 $ vanishes after the first time step as follows from (\ref{eq15}), 
and $\vec {B}$ stays also constant according to (\ref{eq10}), a clear contradiction 
with (\ref{eq8}) arises. Furthermore, equation (\ref{eq9}) predicts that the total 
rotational electric field stays constant, whereas the quasi-static part (\ref{eq14}) 
follows instantaneously all changes of $\vec {B}_0 \left( t \right)$ 
according to (\ref{eq14}).

We note that the quasi-static expressions (\ref{eq7}), (\ref{eq8}), (\ref{eq14}) can be seen as 
solutions of elliptic equations. On the other hand, one obtains from (\ref{eq2}) and 
(\ref{eq4}) by mutual elimination of the fields the inhomogeneous hyperbolic 
equations:
\begin{equation}
\label{eq16}
\Delta \vec {B}-\frac{1}{c^2}\frac{\partial ^2\vec {B}}{\partial 
t^2}=-\frac{4\pi }{c}\,rot\,\vec {j}
\end{equation}
\begin{equation}
\label{eq17}
\Delta \vec {E}_r -\frac{1}{c^2}\frac{\partial ^2\vec {E}_r }{\partial 
t^2}=-\frac{4\pi }{c}\frac{\partial \vec {j}}{\partial 
t}\,+\frac{1}{c^2}\frac{\partial ^2\vec {E}_g }{\partial t^2}\,
\end{equation}
As indicated in [2], the mixture of elliptic and hyperbolic equations 
inherent to Maxwell's system leads apparently to the inconsistencies which 
manifest themselves in the incongruities implied in (\ref{eq10}) as compared to (\ref{eq8}), 
and in (\ref{eq14}) as compared to (\ref{eq9}). The system (1 -- 4) does not permit a 
continuous temporal evolution from a given realistic initial state. In a 
region where the sources in (\ref{eq16}) and (\ref{eq17}) vanish the homogeneous hyperbolic 
equations describe correctly propagating electromagnetic fields, but their 
production mechanism in connection to the sources remains obscure. 

Since in all textbooks it is claimed that Maxwell's equations do have 
solutions which are uniquely determined when the behaviour of the sources is 
given as a function of space and time, we must discuss the usual procedure 
to obtain these solutions which -- according to our analysis -- cannot 
satisfy the first order system. 
\vspace{.6cm}

\noindent \textbf{3. The retarded solutions}

\noindent The normal method of solution expresses the fields by potentials:
\begin{equation}
\label{eq18}
\vec {B}=rot\,\vec {A}\;,\quad \vec {E}=-\nabla \phi 
-\frac{1}{c}\frac{\partial \vec {A}}{\partial t}
\end{equation}
which leads to inhomogeneous wave equations in Lorenz gauge:
\begin{equation}
\label{eq19}
\Delta \phi -\frac{1}{c^2}\frac{\partial ^2\phi }{\partial t^2}=-4\pi \,\rho 
\end{equation}
\begin{equation}
\label{eq20}
\Delta \vec {A}-\frac{1}{c^2}\frac{\partial ^2\vec {A}}{\partial 
t^2}=-\frac{4\pi }{c}\,\,\vec {j}
\end{equation}
They are solved by application of Duhamel's principle to yield the retarded 
solutions, e.g.:
\begin{equation}
\label{eq21}
\vec {A}\left( {\vec {x}{\kern 1pt},\,t} 
\right)=\frac{1}{c}\int\!\!\!\int\!\!\!\int {\vec {j}} \left( {\vec 
{x}'{\kern 1pt},\,t-{\left| {\vec {x}{\kern 1pt}-\vec {x}'{\kern 1pt}} 
\right|} \mathord{\left/ {\vphantom {{\left| {\vec {x}{\kern 1pt}-\vec 
{x}'{\kern 1pt}} \right|} c}} \right. \kern-\nulldelimiterspace} c} 
\right)\,\frac{d^3x'}{\left| {\vec {x}{\kern 1pt}-\vec {x}'{\kern 1pt}} 
\right|}\,\,
\end{equation}

Instead of introducing potentials one may solve the wave equations for the 
fields directly. The magnetic field, for example, can be expressed as the 
sum:
\begin{equation}
\label{eq22}
\vec {B}=\vec {B}_0 +\vec {B}_1 
\end{equation}
where $\vec {B}_0 $ is the instantaneous part (\ref{eq8}), and $\vec {B}_1 $ 
satisfies according to (\ref{eq16}) the equation:
\begin{equation}
\label{eq23}
\Delta \vec {B}_1 -\frac{1}{c^2}\frac{\partial ^2\vec {B}_1 }{\partial 
t^2}=\frac{1}{c^2}\frac{\partial ^2\vec {B}_0 }{\partial t^2}
\end{equation}
In analogy to (\ref{eq21}) this equation has the retarded solution: 
\begin{equation}
\label{eq24}
\vec {B}_1 \left( {\vec {x}{\kern 1pt},\,t} \right)=-\frac{1}{4\pi 
\,c^2}\int\!\!\!\int\!\!\!\int {\left( {\frac{\partial ^2\vec {B}_0 \left( 
{\vec {x}',\,t'} \right)}{\partial t'^2}} \right)} \,\frac{d^3x'}{\left| 
{\vec {x}{\kern 1pt}-\vec {x}'{\kern 1pt}} \right|}\;,\quad t'=t-{\left| 
{\vec {x}{\kern 1pt}-\vec {x}'{\kern 1pt}} \right|} \mathord{\left/ 
{\vphantom {{\left| {\vec {x}{\kern 1pt}-\vec {x}'{\kern 1pt}} \right|} c}} 
\right. \kern-\nulldelimiterspace} c
\end{equation}
Similarly, one may write:
\begin{equation}
\label{eq25}
\vec {E}=\vec {E}_0 +\vec {E}_1 
\end{equation}
and obtain from (\ref{eq17}) a second order differential equation for $\vec {E}_1 $:
\begin{equation}
\label{eq26}
\Delta \vec {E}_1 -\frac{1}{c^2}\frac{\partial ^2\vec {E}_1 }{\partial 
t^2}=\frac{1}{c^2}\frac{\partial ^2\vec {E}_0 }{\partial t^2}\,
\end{equation}
where $\vec {E}_0 $ is the instantaneous part of the electric field 
resulting from (\ref{eq5}) and (\ref{eq12}):
\begin{equation}
\label{eq27}
\vec {E}_0 \left( {\vec {x}{\kern 1pt},\,t} 
\right)=-\frac{1}{c^2}\int\!\!\!\int\!\!\!\int {\left( {\frac{\partial \vec 
{j}\left( {\vec {x}',\,t} \right)}{\partial t}} \right)} 
\,\frac{d^3x'}{\left| {\vec {x}{\kern 1pt}-\vec {x}'{\kern 1pt}} \right|}
\end{equation}
The retarded solution of (\ref{eq26}) is then:
\begin{equation}
\label{eq28}
\vec {E}_1 \left( {\vec {x}{\kern 1pt},\,t} \right)=-\frac{1}{4\pi 
\,c^2}\int\!\!\!\int\!\!\!\int {\left( {\frac{\partial ^2\vec {E}_0 \left( 
{\vec {x}',\,t'} \right)}{\partial t'^2}} \right)} \,\frac{d^3x'}{\left| 
{\vec {x}{\kern 1pt}-\vec {x}'{\kern 1pt}} \right|}\;,\quad t'=t-{\left| 
{\vec {x}{\kern 1pt}-\vec {x}'{\kern 1pt}} \right|} \mathord{\left/ 
{\vphantom {{\left| {\vec {x}{\kern 1pt}-\vec {x}'{\kern 1pt}} \right|} c}} 
\right. \kern-\nulldelimiterspace} c
\end{equation}
It turns out that the fields as obtained from (\ref{eq21}) and (\ref{eq18}) are not the same 
fields as that calculated from (\ref{eq22}), (\ref{eq8}), and (\ref{eq24}), and from (\ref{eq25}), (\ref{eq27}), and 
(\ref{eq28}). This will be demonstrated in the next Section by choosing a specific 
example. Hence, we must conclude that the retarded solutions cannot be 
considered as true solutions of the first order equations.

The reason for this failure must be sought in the inconsistency which lies 
in the fact that equations (\ref{eq20}), (\ref{eq23}), (\ref{eq26}) connect the sources $\vec 
{j}\left( {\vec {x}{\kern 1pt},t} \right)$, $\vec {B}_0 \left( {\vec 
{x}{\kern 1pt},t} \right)$, $\vec {E}_0 \left( {\vec {x}{\kern 1pt},t} 
\right)$, respectively, with the travelling wave fields $\vec {A}\left( 
{\vec {x}{\kern 1pt},t} \right)$, $\vec {B}_1 \left( {\vec {x}{\kern 1pt},t} 
\right)$, $\vec {E}_1 \left( {\vec {x}{\kern 1pt},t} \right)$ at the \textit{same }time 
$t$, whereas in the retarded solutions (\ref{eq21}), (\ref{eq24}), (\ref{eq28}) the differentiation 
of the source is dated back to the earlier time $t'=t-{\left| {\vec 
{x}{\kern 1pt}-\vec {x}'{\kern 1pt}} \right|} \mathord{\left/ {\vphantom 
{{\left| {\vec {x}{\kern 1pt}-\vec {x}'{\kern 1pt}} \right|} c}} \right. 
\kern-\nulldelimiterspace} c$. As pointed out in [2], the source may be very 
far away from the observation point, and may not even exist anymore when the 
fields $\vec {B}_1 \left( {\vec {x}{\kern 1pt},t} \right)$, $\vec {E}_1 
\left( {\vec {x}{\kern 1pt},t} \right)$ are measured at time $t$. It makes 
little sense to differentiate non-existent instantaneous fields at time $t$, 
but this was necessary to derive equations (\ref{eq16}), (\ref{eq17}) from the system (1 -- 
4). Obviously, it constitutes a \textit{contradictio in adjecto }connecting the travelling fields predicted 
by (\ref{eq16}) and (\ref{eq17}) with the stationary sources in the first order system at 
the same time.
\vspace{.6cm}

\noindent \textbf{4. A specific example}

\noindent In order to facilitate the calculations we choose an example where we have 
$div\,\vec {j}=0$. In this case the scalar potential vanishes because of 
$\rho =0$ which makes Lorenz and Coulomb gauge identical: $div\,\vec {A}=0$. 
The chosen example is a hollow cylinder which carries a closed oscillating 
current driven by an rf-generator through a resistor R, as sketched in Fig. 
1. It is assumed that the current was switched on at time $t=-\infty $ and 
oscillates with a sinusoidal time dependence: $I\,\exp \left( {-i\,\omega 
\,t} \right)$. The current flows in a thin central filament, and returns 
symmetrically on the cylindrical surface. This can be achieved to an 
arbitrary degree of accuracy, if the inverse wave vector $k=\omega 
\mathord{\left/ {\vphantom {\omega c}} \right. \kern-\nulldelimiterspace} c$ 
is large compared to the dimensions of the device.

\begin{figure}[htbp]
\centerline{\includegraphics[width=3.in,height=4.in]{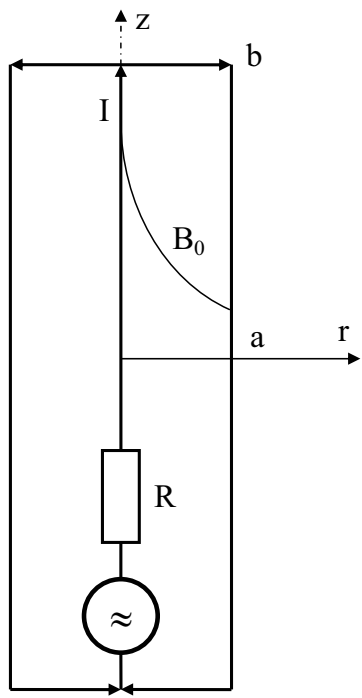}}
\label{fig1}
\caption{\large Oscillating current flowing in a closed circuit of cylindrical 
geometry}
\end{figure}
\newpage
\noindent The instantaneous magnetic field component (\ref{eq8}) produced in this configuration is:
\begin{eqnarray}
\label{eq29}
B_0 &=&\frac{2\,I}{c\,r}\exp \left( {-i{\kern 1pt}\omega {\kern 1pt}t} 
\right)\;,\quad r\le a\;,\quad -b\le z\le b \\ 
 B_0 &=&0\;,\quad r>a\;,\quad z<-b\;,\quad z>b \nonumber
 \end{eqnarray}
and the instantaneous electric field (\ref{eq27}) becomes: 
\begin{eqnarray}
\label{eq30}
\vec {E}_0 =\frac{i\,k\,I}{2\pi \,c}\,\exp \left( {-i{\kern 1pt}\omega 
{\kern 1pt}t} \right)\int\limits_0^{2\pi } {\left\{ {\int\limits_0^a {\left[ 
{\frac{\cos \varphi '\,dr'}{R}} \right]_{z'=-b}^{z'=+b} \,\vec {e}_r 
+\int\limits_{-b}^b {\left[ {\frac{dz'}{R}} \right]_{r'=a}^{r'=0} \vec {e}_z 
} } } \right\}} \,d\varphi ' \\ 
 R=\sqrt 
{r^2+r'^2-2\,r\,r'\cos \varphi '+\left( {z-z'} \right)^2} \nonumber 
 \end{eqnarray}
The retarded solution of the vector potential as obtained from (\ref{eq21}) is:
\begin{equation}
\label{eq31}
\vec {A}=\frac{I\,e^{-i{\kern 1pt}\omega {\kern 1pt}t}}{2\pi 
\,c}\int\limits_0^{2\pi } {\left\{ {\int\limits_0^a {\left[ {\frac{\exp 
\left( {i\,k\,R} \right)}{R}} \right]} _{z'=-b}^{z'=+b} \cos \varphi 
'\,dr'\,\vec {e}_r +\int\limits_{-b}^{+b} {\left[ {\frac{\exp \left( 
{i\,k\,R} \right)}{R}} \right]_{r'=a}^{r'=0} dz'\,\vec {e}_z } } \right\}} 
\,d\varphi '
\end{equation}
It may be substituted into (\ref{eq18}) to yield the fields as given by Jackson for 
a localized oscillating source [4]:
\begin{eqnarray}
\label{eq32} 
B\!\!&=&\!\!\frac{I\,e^{-i{\kern 1pt}\omega {\kern 1pt}t}}{2\pi 
\,c} \int\limits_{0}^{2\pi}\left\{\int\limits_0^a {\left[ 
{\frac{e^{i\,k\,R}\left( {1-i\,k\,R} \right) \,s\,\cos \varphi ' dr'}{R^3}} 
\right]_{s=z-b}^{s=z+b} } \right. \nonumber \\ &\quad& \quad \quad \quad 
\left.-\int\limits_{z-b}^{z+b} {\left[ 
{\frac{e^{i\,k\,R}\left( {1-i\,k\,R} \right)\,\left( {r-r'\cos \varphi '} 
\right)\,dz'}{R^3}} \right]_{r'=0}^{r'=a} } \right\} \,d\varphi '
\end{eqnarray} 

\begin{equation}
\label{eq33}
\vec {E}=\frac{i\,k\,I\,e^{-i{\kern 1pt}\omega {\kern 1pt}t}}{2\pi 
\,c}\int\limits_0^{2\pi } {\left\{ {\int\limits_0^a {\left[ 
{\frac{e^{i\,k\,R}\cos \varphi '\,dr'}{R}} \right]_{z'=-b}^{z'=+b} \,\vec 
{e}_r +\int\limits_{-b}^b {\left[ {\frac{e^{i\,k\,R}\,dz'}{R}} 
\right]_{r'=a}^{r'=0} \vec {e}_z } } } \right\}} \,d\varphi '
\end{equation}
where $s=z-z'$. It is doubtful whether these solutions satisfy also the 
differential equations (\ref{eq23}) and (\ref{eq26}). In order to check on this we consider, 
e.g., equation (\ref{eq23}) adapted to our case:
\begin{equation}
\label{eq34}
r^2\frac{\partial ^2B_1 }{\partial r^2}+r\,\frac{\partial B_1 }{\partial 
r}-B_1 \left( {1-r^2k^2} \right)+r^2\frac{\partial ^2B_1 }{\partial 
z^2}=-\frac{2\,k^2I\,r}{c}\,e^{-i\,\omega \,t}
\end{equation}
where the right-hand-side must be set to zero outside the cylinder of Fig. 
1. We integrate this equation with respect to $r$ and obtain:
\begin{equation}
\label{eq35}
\frac{\partial B_1 }{\partial r}-\frac{B_1 }{r}+\frac{1}{r^2}\int\limits_0^r 
{r^2} \left( {\frac{\partial ^2B_1 }{\partial z^2}+k^2B_1 } 
\right)\,dr=-\frac{\,k^2I}{c}\,e^{-i\,\omega \,t}
\end{equation}
The contribution $B_1 $ may be calculated from (32) by expansion of the 
exponential function for $k\,R<1$. In zero order one obtains the 
instantaneous field (\ref{eq29}), and in second order one has:
\begin{equation}
\label{eq36}
B_1 =-\frac{I\,k^2e^{-i{\kern 1pt}\omega {\kern 1pt}t}}{4\pi 
\,c}\int\limits_0^{2\pi } {\left\{ {\int\limits_0^a {\left[ {\frac{s\cos 
\varphi 'dr'}{R}} \right]_{s=z-b}^{s=z+b}\!\!\!\! -\int\limits_{z-b}^{z+b} {\left[ 
{\frac{\left( {r-r'\cos \varphi '} \right)\,dz'}{R}} \right]_{r'=0}^{r'=a} } 
} } \right\}d\varphi ' }+ O\left( {k^{n>2}} \right)
\end{equation}
The integration over $r'$ and $z'$ may be carried out analytically to yield:
\begin{equation}
\label{eq37}
B_1 =\frac{I\,k^2e^{-i{\kern 1pt}\omega {\kern 1pt}t}}{4\pi 
\,c}\int\limits_0^{2\pi } {\left[ {s\cos \varphi '\ln \left( {r'-r\cos 
\varphi '+R} \right)+\left( {r-r'\cos \varphi '} \right)\ln \left( {s+R} 
\right)} \right]_{s=z-b,\,r'=0}^{s=z+b,\,r'=a} d\varphi '} 
\end{equation}
Expanding this expression in a power series of $r$, and inserting it into 
the left-hand-side of (\ref{eq34}) we find for $z=0$:
\begin{equation}
\label{eq38}
\frac{I\,k^2e^{-i{\kern 1pt}\omega {\kern 1pt}t}}{2\,c}\,\left[ {\left( 
{\frac{b}{\left( {a^2+b^2} \right)^{\frac{3}{2}}}-\frac{1}{b^2}} 
\right)\,r^2+\left( {\frac{1}{2\,b^4}-\frac{b\left( {2b^2-3a^2} 
\right)}{4\left( {a^2+b^2} \right)^{\frac{7}{2}}}} \right)\,r^4} 
\right]\;+O\left( {r^{n>4}} \right)
\end{equation}
which is obviously at variance with the right-hand-side of (\ref{eq35}). A similar 
conclusion is reached, if (\ref{eq33}) is substituted into (\ref{eq26}). This can only be 
checked numerically, since the instantaneous field $\vec {E}_0 $ does not 
vanish outside the cylinder, in contrast to $\vec {B}_0 $.

Result (\ref{eq38}) proves that the standard solutions (\ref{eq32}) and (\ref{eq33}) do not satisfy 
the first order system from which equations (\ref{eq16}) and (\ref{eq17}) were derived. 
Hence, our conclusion in Sect. 2., namely that the first order system does 
not permit a solution, cannot be refuted by referring to the retarded 
solutions as taught in the textbooks such as [4]. 

There is also a physical reason to reject Jackson's solution (\ref{eq31}) for the 
considered case. If one calculates the fields with (\ref{eq18}) from (\ref{eq31}) and 
evaluates the Poynting vector $\vec {E}\times \vec {B}$ at large distance, 
one can integrate the total radiation power emitted by the closed circuit of 
Fig. 1:
\begin{equation}
\label{eq39}
P_{tot} =\int\!\!\!\int {\frac{c}{4\pi }} \left( {\vec {E}\times \vec {B}} 
\right)\cdot \vec {d}^2x=\frac{I^2\,a^4\,b^2\,k^6}{6\,c}
\end{equation}
This result is obviously not physical. The device in question may be seen as 
a short-circuited cable which should not continuously loose energy to the 
outside world; in particular not when the enclosing shell would be made out 
of superconducting material. The predicted power loss (\ref{eq39}) could certainly 
not be confirmed experimentally.

\vspace{.6cm}

\noindent \textbf{5. Conclusions}

\noindent It has been shown that an attempt to calculate numerically the temporal 
evolution of the electromagnetic field from the full set of Maxwell's first 
order equations will fail due to the internal inconsistencies built into the 
coupled system of equations. As noted earlier [2], the reason lies in the 
fact that the travelling wave fields are connected with the stationary 
sources at the same time.

Maxwell's equations describe correctly the production of the instantaneous 
electromagnetic field, and also the propagation of wave fields in empty 
space. The production mechanism of electromagnetic waves by time varying 
sources, however, does not find an explanation in the framework of Maxwell's 
theory. Contrary to what is commonly believed, the retarded solutions for 
the electromagnetic potentials do not lead to fields which are in agreement 
with a direct solution of the second order differential equations for the 
fields.

\vspace{.6cm}
\newpage
\noindent \textbf{Acknowledgment}

\noindent The author is indebted to V. Onoochin for initiating this work. Vladimir 
contributed significantly in the early discussions, but he modestly felt 
that he should not be a co-author of the paper.

\end{large}

\end{document}